# Evaluating dispersion models for *ab initio* simulation of Group-I and Group-II molten fluoride salts

Shubhojit Banerjee[a*], Rajni Chahal-Crockett[b], Julián Barra[a], Stephen T Lam[a*]

[a] University of Massachusetts Lowell, 1 University Ave, Lowell, MA-01854

[b] Mechanical and Nuclear Engineering, Tennessee Technological University, Cookeville, TN-38505

**AUTHOR INFORMATION**

**Corresponding Authors**:

*Shubhojit Banerjee: shubhojit_banerjee@student.uml.edu

* Stephen T Lam: stephen_Lam@uml.edu

## Abstract

Ab initio molecular dynamics (AIMD) based on density functional theory (DFT) is a powerful approach for modeling molten salts. However, standard exchange-correlation functionals often neglect dispersion interactions, introducing potential errors in property predictions. Dispersion corrections are commonly applied ad hoc to match experimental salt densities, but their systematic impact on predicting structure, thermophysical, and transport properties of salt remains unexamined. This study evaluates the impact of Grimme's DFT-D and nonlocal van der Waals (vdW-DF) corrections on molten fluorides of Group I (LiF, NaF, KF) and Group II ($BeF_2$, $MgF_2$, $CaF_2$), which are relevant to reactor applications. Results indicate that dispersion corrections have a minor effect on binding energies but significantly influence density predictions. Systematic benchmarking across compositions and temperatures reveals that semi-empirical dispersion models often produce more accurate densities compared to vdW-DF. Diffusion coefficients remain largely invariant to dispersion corrections at fixed densities, while coordination number distributions exhibit notable differences based on chosen dispersion. $BeF_2$, in particular, deviates from other fluorides, showing pronounced structural and dynamical differences in the absence of dispersion corrections. This highlights the necessity of dispersion effects for high-charge-density cations that promote intermediate- to long-range ordering. These findings provide a systematic framework for selecting dispersion models in molten salt simulations, improving density and structural predictions.

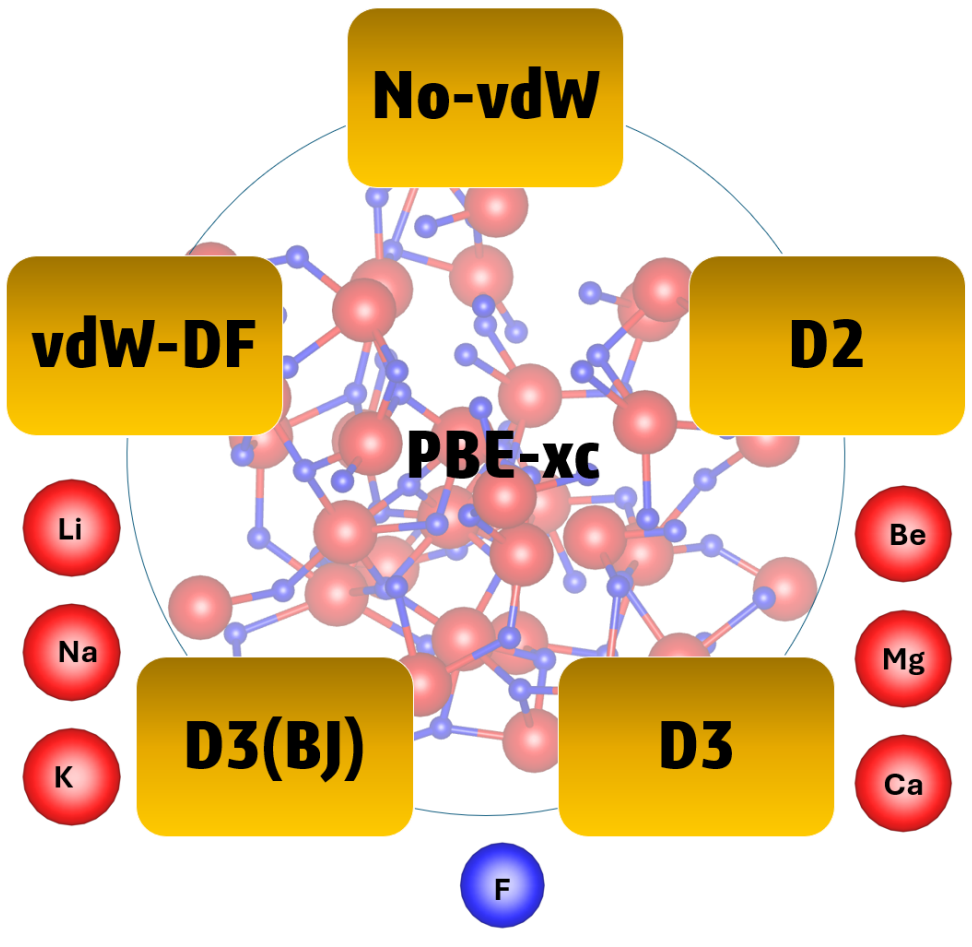

## 1. INTRODUCTION

Molten salts are a promising class of materials in a wide range of applications, including solar thermal energy storage, next-generation nuclear energy systems, and advanced batteries, owing to their ability to serve as effective heat transfer fluids under high-temperature, ambient-pressure, and high-radiation-flux conditions.[1–5] However, the experimental measurement of the thermophysical and structural properties of these salts is extremely challenging due to their extremely harsh operating conditions and costly experimental setups. Herein, the utility of computer simulations has been demonstrated to be highly effective and complementary to experiments by allowing improved interpretation and enhancement of experimental observations.[6]

One common practice is parameterized interatomic potentials based classical molecular dynamics, which are often used to predict the structure, dynamics, and thermophysical properties of molten salts.[2,7] However, their performance is questionable when they are used to study structures that differ significantly from those on which the parameters were fitted. Aside from this, limitations in the functional form of the selected potential could lead to inaccuracies and erroneous results.[2] One way to overcome this challenge is by using neural network inter-atomic potentials (NNIP)[8,9], which are known for their accuracy and scalability.[10] However, the success of these NNIPs relies on the accuracy of the training data[10,11], which is often chosen from *ab initio* molecular dynamics (AIMD). AIMD is parameter-free, where the atomic forces are directly computed with density functional theory (DFT), providing a favorable cost-to-performance ratio. The accuracy of AIMD simulations crucially hinges on the choice of basis set and exchange-correlation functionals used. The plane wave basis sets with local density approximation (LDA) or generalized gradient approximation (GGA)[12] exchange-correlation functionals demonstrate promising performance for periodic solid and liquid systems.[13–16] However, these conventional semi-local exchange-correlation (xc) functionals, often fall short of adequately capturing nonlocal correlations, namely dispersion interactions.

As such, various computationally affordable dispersion models have been proposed to account for dispersion interactions and have been widely adopted to improve their description in diverse systems. One such model is Grimme's proposed density functional theory with dispersion corrections (DFT-D),[17,18] which incorporates semi-empirical correction terms derived from interatomic potentials to enhance the treatment of long-range interactions. Subsequently, Dion et al.[19] introduced the nonlocal van der Waals density functional (vdW-DF) method, which computes nonlocal correlation energy without relying on empirical considerations.

Following the development of dispersion models, their importance was tested on different systems. For example, in water, it was found that the absence of dispersion interaction underestimated the density by almost 20%.[20] Similarly, dispersion corrections have been shown to impact the structural and dynamic properties of room-temperature ionic liquids and other non-covalent ionic systems.[21,22] Furthermore, comparative studies have also revealed that selecting an optimal dispersion model is critical for ensuring accuracy, emphasizing the need for careful selection of dispersion models.[23–25]

For molten salts as well, Dario et al. have shown the absence of a dispersion effect in density and the melting point of LiF. Lam et. al. showed that neural network interatomic potentials trained for the FLiBe system produce ~14% error in density as the training AIMD data lacks vdW interactions.[10,11]Further Nam et al. have shown that the inclusion of vdW interaction improves the density prediction of FLiBe. In a recent study, Rajni et.al have shown the effect of different dispersion models in accurately predicting the structure and density of $AlCl_3$.[7]

As such, previous studies have examined dispersion corrections for individual salts—with a focused on predicting density. The effects of dispersion corrections on a broader range of molten salt properties, including structure, thermophysical behavior, and transport properties, have not been systematically analyzed. Furthermore, a comprehensive comparison across a variety of molten salt compositions relevant to reactor design is lacking. This is particularly important since the choice of dispersion model, along with the potential errors arising from omitting or misapplying (ad hoc choice of dispersion) these corrections, can have significant implications for predicting salt behavior in reactor environments, where operating conditions can vary widely (such as wide range of temperature fluctuation, the incursion of corrosion product, etc.). Herein, a systematic investigation comparing different dispersion models across various salt systems to find the optimal model would be critical to successfully developing predictive models for the molten salts.

To bridge this gap, we performed AIMD simulations using PBE functional with different dispersion corrections for the molten fluoride salts involving Group-I ($Li^{+}$, $Na^{+}$, $K^{+}$) and Group-II ($Be^{2+}$, $Mg^{2+}$, $Ca^{2+}$) cation species (a common constituent of reactor-relevant salts). For each system, a case without dispersion corrections (a.k.a, no-vdW) was also considered. In specific, we systematically evaluated semiempirical dispersion correction methods, namely Grimme's DFT-D2[26], DFT-D3[17], DFT-D3(BJ)[27], and a nonlocal vdW-DF functional[19]. For all cases, to assess the impact of no dispersion as well as different dispersion methods, we compared the predicted structure as well as properties of these molten salts with the reported experimental values (where available). Specifically, to evaluate the local salt structure, our analysis remains focused on the radial distribution functions and coordination environments, while thermodynamic properties (density, binding energy) and transport properties (self-diffusion coefficients) were considered to evaluate the accuracy of dispersion interactions for each molten salt. We found that while dispersion minimally affects cation-anion binding energies, it significantly influences density predictions. As such, calculations with dispersion correction consistently overpredict density, whereas in the absence of dispersion, density gets underpredicted. Despite these differences, structural properties and self-diffusion coefficients remain largely consistent across dispersion models when simulations are performed at experimental densities. Notably, for high charge density cations like $Be^{2+}$, we found that the choice of dispersion correction method plays a crucial role in accurately capturing local structure and dynamics. As such, the results reported in this study guide the selection of dispersion models in AIMD simulations of different high-temperature salt systems.

## 2. DISPERSION CORRELATIONS

There are several dispersion models available for AIMD simulations. However, in the current work, we used four different dispersion correlations, that is, DFT-D2[26], DFT-D3[17], DFT-D3(BJ)[27], and the nonlocal vdW-DF functional[19]. The choice of these dispersion models are motivated by their successful applications for different molten salt systems. In general, the DFT-D method incorporates a semi-empirical correlation to the energy obtained from DFT calculations to account for dispersion interactions. The total energy taking dispersion into account is given by Eq 1.

$$E_{DFT-D} = E_{KS-DFT} + E_D^{disp} \tag{1}$$

Here, $E_{KS-DFT}$ represents the energy calculated using the Kohn-Sham equations for a selected exchange-correlation functional, and $E_D^{disp}$ accounts for the semi-local dispersion energies. For instance, DFT-D2, as developed by Grimme, adds a semi-empirical dispersion correction to standard DFT, expressed as:

$$E_{disp}^{D2} = -S_6 \sum_{\{i>j\}} \frac{C_6^{ij}}{r_{ij}^6} f(r_{ij}) \tag{2}$$

In Eq. 2, $C_6^{ij}$ is the dispersion coefficient, $r_{ij}$ is the distance between atoms, and $f(r_{ij})$ is a damping function. While the $C_6$ coefficient for the D2 method is a constant that depends on polarization and ionization energy of the chosen atomic pairs, DFT-D3 refines this by introducing environment-dependent $C_6^{ij}$ and higher-order dispersion terms ($C_8^{ij}$). The empirical form of D3 dispersion is shown in Eq. 3:

$$E_{disp}^{D3} = -\sum_{\{i<j\}} \left[ \frac{C_6^{ij}}{r_{ij}^6} + \frac{C_8^{ij}}{r_{ij}^8} \right] f(r_{ij}) \tag{3}$$

Where $C_8^{ij}$ is the higher-order dispersion coefficient and $f_{D3}(r_{ij})$ is the damping function. There are two different flavors of the D3 model available based on the use of the damping function. While the original one uses a damping function by Head-Gordon, a modified version, namely D3(BJ) modifies the damping function using the Becke-Johnson (BJ) damping[27]. Both D3 and D3(BJ) don't affect long-distance dispersion effects. However, the use of BJ damping provides a constant dispersion effect at the short atomic distance, whereas the Head-Gordon dispersion goes to zero at short distances, leading to a unphysical scenario.[27] Previous work[27] has showed the D3 method can lead to longer interatomic distances with dispersion correction due to this repulsive effect, which can be overcome with BJ damping.

In contrast to semiempirical dispersion models, van der Waals density functionals (vdW-DFs)[19] include a nonlocal correction term incorporated into the exchange-correlation functional (Eq. 4)

$$E_c^{nl} = \frac{1}{2} - \sum_{\{i<j\}} \int \rho_i(\boldsymbol{r})\Phi(\boldsymbol{r},\boldsymbol{r}'),\rho_j(\boldsymbol{r}')\,dr\,dr' \tag{4}$$

Here, $\rho_i(\boldsymbol{r})$, $\rho_j(\boldsymbol{r})$are the electron densities of atoms $i$ and $j$, and $\rho_i(\boldsymbol{r})\Phi(\boldsymbol{r},\boldsymbol{r}'),\rho_j(\boldsymbol{r}')$ is a nonlocal kernel that depends on the electron density at both positions. While DFT-D requires the fitting of the parameter, nonlocal vdW-DFs are parameter-free, ensuring better transferability. However, DFs are computationally expensive compared to the DFT-D method.

## 3. SIMULATION PROTOCOL

The first principles molecular dynamics simulations are carried out with the Vienna *ab initio* Simulation Package (VASP).[28] A plane wave basis set with projector augmented wave (PAW) method and Perdew-Burke-Ernzerhof (PBE) GGA[12] exchange-correlation functional was used for underlying density functional theory (DFT) calculations. While the local density approximation (LDA) is frequently employed in electronic structure calculations, the GGA functional is chosen here due to its improved accuracy in capturing gradient-dependent corrections to the exchange-correlation energy, which are particularly relevant for describing intermolecular interactions with minimal additional computational cost. For sampling reciprocal space, a 1 ×1 × 1 gamma-centered k-point mesh was used. The DFT parameters are chosen to yield an energy convergence within 2 meV/atom. A plane wave basis set with a cutoff of 650 eV, along with a convergence criterion of $10^{-5}$ is chosen for self-consistent field (SCF) calculations. The chosen pseudo potentials correspond to those recommended in the VASP documentation. All the calculations are carried out using an NVT ensemble with a Nosé-Hoover thermostat, at experimental density computed at different temperatures. The simulation temperatures (Table 1) were set at least ~10 K above the salt's melting point to ensure that, at the chosen density, the salt remains well in the liquid phase. The initial structures are generated by randomizing their positions with packmol.[29]

**Table 1.** Simulation details, including different simulation temperatures and their corresponding box lengths. The melting point of the simulated molten salts are also included for reference. Simulations chosen for structural comparison are mentioned in bold.

| | **Melting point $T_m$ (K)** | **Simulation temperature (K) (Box length)** | | |
|---|---|---|---|---|
| **LiF** | 1121.2 | 1250 (9.46) | **1300 (9.51)** | 1350 (9.56) |
| **NaF** | 1268 | **1300 (10.74)** | 1337 (10.786) | 1373 (10.829) |
| **KF** | 1131.2 | 1200 (12.287) | 1250 (12.36) | **1300 (12.4)** |
| **$BeF_2$** | 821.2 | 1073 (11.125) | **1100 (11.125)** | 1123 (11.126) |
| **$MgF_2$** | 1534.2 | 1700 (11.606) | 1850 (11.739) | **2000 (11.878)** |
| **$CaF_2$** | 1692 | 1700 (12.137) | 1950 (12.2973) | **2200 (12.465)** |

As previously mentioned, all the calculations were carried out both with and without dispersion correlations. All calculations with semi-empirical dispersions or without dispersions were carried out for almost 80 ps with a time step of 2 fs. The first 40 ps are run for equilibration, and the rest are used for evaluating the properties of molten salts. The simulations with vdW-DF dispersion were run for 60 ps, from which at least last 30 ps were used for property calculations (shorter simulations are due to their high computational cost). While different temperature calculations have been used for density and diffusion coefficient calculations, a single temperature has been used for structural comparison, which is mentioned in Table 1 in bold text.

## 4. RESULTS AND DISCUSSION

To evaluate the performance of different dispersion models, we focused on their ability to predict key properties such as salt density, transport properties like self-diffusion coefficients, and local structural features. For both structure and properties, experimental data (where available) for each salt is used to evaluate the accuracy of the predictions in each case. More details will be discussed in detail in the following sections.

### 4.1. Excess pressure and density

Accurate density determination is crucial for molten salts, as it directly influences composition, local structures, and key transport properties like diffusivity, viscosity, and thermal conductivity, ensuring meaningful comparisons with experimental data.[30,31] To elucidate the impact of considered dispersion interactions on density, Nam et al. previously employed the third-order Birch-Murnaghan equation of state (EOS) fitting method to compute simulated volumes explicitly.[24] This approach necessitates numerous NVT simulations at varied volumes, which are computationally intensive and unsuitable for efficiently evaluating dispersion functionals across a wide range of systems. For this reason, instead of exploring the exact simulated density for individual systems, we estimate the density error by performing calculations of excess pressures for single NVT simulations for each molten salt at their experimental densities taken from the Molten Salts Thermal Properties Database (MSTDB).[32]

In this approach, the total pressure exerted on the system serves as an indicator of density. Ideally, if the simulated density matches the experimental density, the system should exhibit zero total pressure. However, the AIMD-simulated density deviates from the experimental density, leading to either positive or negative pressure in the simulation cell. This pressure deviation provides insight into the direction and magnitude of density error, as described by the relationship (Eq. 5)

$$\frac{\partial \rho}{\rho} = -\beta \partial p \tag{5}$$

Where $\beta$ is the isothermal compressibility represents the relative density error, and $\partial$p is the deviation of total pressure from zero. We use AIMD-calculated total pressure as $\partial$p and employ experimentally measured $\beta$ values from Marcus[33], which are reported at 1.1$T_m$ ($T_m$= melting point) (Table S1). While β may vary slightly with temperature, this variation is expected to be modest between the measured and simulated temperatures.[34] Since our goal is to determine relative density errors across different dispersion models and chemical systems, rather than exact density values, this method provides a practical and consistent way to compare dispersion models.

In this regard, a positive cell pressure indicates that the equilibrium volume from the simulation would be overpredicted (if the cell is allowed to relax), and consequently, the density would be underpredicted. In contrast, negative cell pressure indicates an overprediction of density. Overall, this method not only expedites the analysis process but also facilitates a comprehensive assessment of density over- or under-prediction, enabling a faster evaluation of appropriate dispersion models for molten salt systems under investigation.

To verify the statistical reliability of the calculated pressures, we monitored the instantaneous pressure across the equilibrated trajectory. In all the cases, the pressure fluctuations consistently showed stability around a mean value with standard deviations of approximately ±2-3 kB, confirming thermodynamic equilibrium. Detailed pressure stability plots are provided in the Supporting Information (Figures S1-S6). Figure 1 presents excess average pressures for Group I (G-I) and Group II (G-II) fluoride salts across multiple temperatures and dispersion corrections.

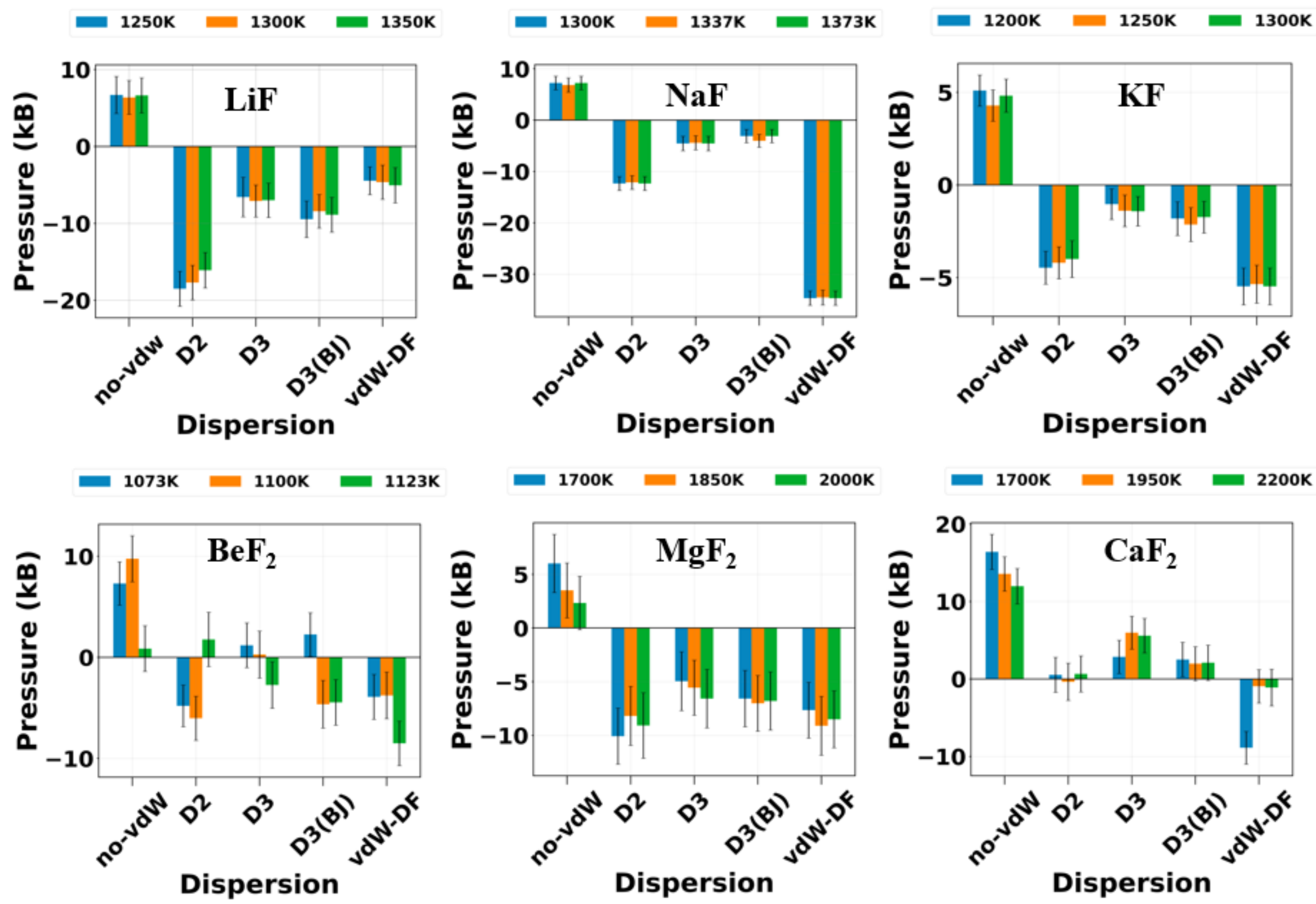

**Figure 1.** Excess pressure on the G-I (top) and G-II (bottom) molten salts obtained from NVT simulations.

Without the use of a dispersion correction (no-vdW), the cell pressures are consistently positive, with values ranging between 5 and 15 kB, indicating density underestimation. Here, pressures are generally lower for Group I metals, exhibiting slightly lower pressures, with LiF,

NaF, and KF exhibiting maximum pressures of ~7 kB, 7 kB, and 5 kB, respectively. For Group I, the excess pressure remains roughly constant (within uncertainty) between homologous temperatures (T/Tm) of 1.0 to 1.2 for a given salt. Meanwhile, although Group II metal fluorides exhibited positive pressures, the magnitudes tended to be higher compared to Group I, with maximum pressures of 10 kB, 5 kB, and 15 kB for BeF2, MgF2, and CaF2, respectively. Furthermore, significant temperature dependence was observed between T/Tm of 1.0 to 1.2, where pressures systematically decreased with an increase in temperature for MgF2 and CaF2. This is potentially caused by an increase in dynamic disorder, which has a larger effect on determining density compared to dispersion at higher temperatures. However, this systematic temperature effect is not observed for BeF2, where small changes in T result in large non-monotonic changes in pressure from 10 kB at 1100 K to 1123 K. This could be attributed to long-range polymeric structures that are observed, as discussed in ref[7,35].

The semi-empirical D2 correction consistently yields negative pressures across nearly all salts and temperatures (ranging from -5 to -20 kB for Group I and 0 to -10 kB for Group II), which indicates the density overprediction. The D2 method applies a pairwise dispersion correction that lacks an environment-dependent component, which can lead to an overestimation of attractive dispersion forces, thereby lowering the cell pressure. As temperature increases, thermal motion partially counteracts this effect, and pressures gradually decrease, yet the absence of environment-specific adjustments keeps the D2-induced pressure reduction significant.

In contrast, the D3 and D3(BJ) corrections include environment-dependent terms, which most likely mitigate the overestimation of the attractive effect seen in D2, leading to pressure values lower than those of D2, although the system pressure remains in the negative regime. This indicates improvement in density prediction over D2, while the density is still overpredicted. Specifically, for Group I fluorides, these models produce pressures in the range of -1 to -5 kB, with NaF and KF showing slightly less negative pressures compared to LiF. For Group II fluorides, D3 and D3(BJ) reduce pressure magnitudes to between -5 kB and 5 kB ($CaF_2$). Moreover, the pressure remains consistent across temperatures for both G-I and G-II (within the uncertainty).

The non-local vdW-DF functional generally resulted in negative pressures across most salts and temperatures (ranging from -1 to -30 kB), indicating overestimated densities. LiF and KF exhibited moderate negative pressures (~-5 kB), while NaF showed a substantially larger pressure of approximately -30 kB. G-I salts overall demonstrated relatively consistent pressure responses across the simulated temperatures. However, Group-II salts showed more pronounced temperature dependence. For example, $BeF_2$ exhibited a pressure drop from -5 to -10 kB between 1100 K and 1123 K, potentially due to the disruption of extended polymeric structures. In contrast, $CaF_2$ showed a pressure increase from -9 to -1 kB when the temperature increased above 1700 K. This behavior is likely related to the proximity of 1700 K to the melting point of $CaF_2$. For $MgF_2$ excess pressure remains consistent across temperatures. These results indicate that the predictive behavior of vdW-DF is strongly system- and temperature-dependent.

To further quantify density errors derived from excess pressures, we have used Eq. 5, and the results are shown in Figure S7. Empirical corrections D2 overpredict density by 0-20 %. Whereas D3, D3(BJ) typically reduce the overestimation of densities and produce errors between 2-5%. The vdW-DF exhibited greater variability, with density overestimations ranging from 1% to as high as 45%, depending on the salt and temperature. Notably, this general behavior is consistent with prior studies. For example, in a study of the KCl–NaCl–$MgCl_2$ system[36], the absence of dispersion corrections led to density underestimation by ~16%, while D2, D3, and D3(BJ) overestimated density by approximately 8%. Similarly, in FLiBe[24], D2 was found to overpredict density, while PBE (without dispersion) underestimated it. These results support the broader conclusion that dispersion corrections tend to over-predict density, whereas in the absence of dispersion correction, the density becomes underpredicted.

Comparison across individual systems shows that for Group-I salts, D3(BJ) and D3 give the most accurate predictions for NaF and KF, respectively, resulting in density errors of approximately 2.5%. For LiF, vdW-DF provides the lowest pressure across temperatures, corresponding to a density overestimation of ~5%. For Group-II salts, the most suitable model varies by composition. D3 yields the lowest density error for $BeF_2$ and $MgF_2$. In contrast, D2 gives the most accurate density for $CaF_2$, with an error close to 1%. Given the challenges associated with high-temperature density measurements, a density error within 0–5% is considered acceptable and likely reflects a combination of simulation limitations, thermodynamic fluctuations, and experimental uncertainties.

Overall, our results demonstrate that empirical dispersion models, particularly D3 and D3(BJ), tend to outperform non-local vdW-based corrections for fluoride salts in most cases. However, the choice of dispersion model should be guided by both system characteristics and validation against experimental benchmarks to ensure reliable thermophysical property predictions in molten salt simulations.

### 4.2. Salt structure analysis

In the previous section, we have seen the effect of different dispersion models on salt density. Herein, we will see the impact of different dispersions on the salt structure. Specifically, we analyze radial distribution functions (RDF), coordination numbers, and angular distribution functions (ADF) computed with different dispersion models across the selected salts. The detailed observations are presented in the following sections.

### *4.2.1. Radial distribution function*

To study the effect of different dispersion models on G-I and G-II metal fluorides' structure, the radial distribution functions (RDF) for cation-fluorine (M-F), fluorine-fluorine (F-F), and cation-cation (M-M) are computed and compared in Figure 2 and Figures S14-S15. A structural convergence test is first performed based on the block averaging method, and the results are shown in Figures S8-S13. The low standard error on the average RDF enhances the reliability of the results.

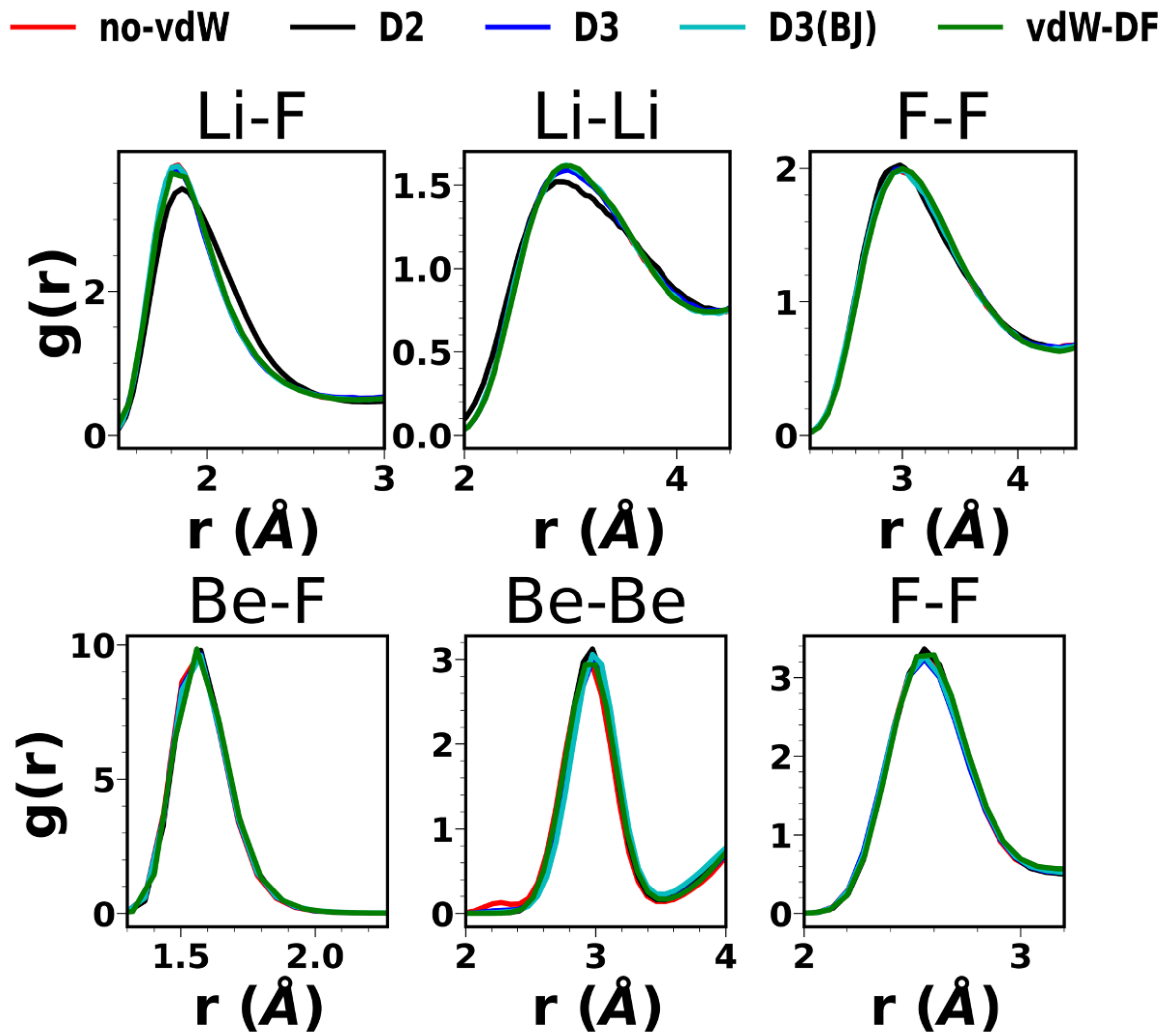

**Figure 2**. Partial radial distribution functions (RDF) for cation-anion, cation-cation, and anion-anion for LiF (upper panel) and BeF2 (lower panel) molten salts with and without different dispersion models.

The key features of RDFs, namely M-F's first peak position (average cation-anion bond length) and the dimension of the solvation shell (first minimum in M-F RDF) for all the G-I and G-II salts are shown in Tables 2a and 2b.

**Table 2a**. Dimension of the solvation shell (first minima in M-F PDF) for G-I and G-II salts across different dispersion models.

| | **No-vdW** | **D2** | **D3** | **D3(BJ)** | **vdW-DF** |
|---|---|---|---|---|---|
| **LiF** | 2.852 | 2.9 | 2.78 | 2.8523 | 2.84 |
| **NaF** | 3.54 | 3.5 | 3.5 | 3.46 | 3.45 |
| **KF** | 4.025 | 4.095 | 3.885 | 3.955 | 4.05 |
| **$BeF_2$** | 2.275 | 2.275 | 2.275 | 2.275 | 2.28 |
| **$MgF_2$** | 2.94 | 2.975 | 2.835 | 2.905 | 2.92 |
| **$CaF_2$** | 3.395 | 3.325 | 3.395 | 3.395 | 3.4 |

For both the G-I and G-II salts, the first solvation shell dimension remains largely unchanged across different dispersion models, indicating that the local environment (short-range ordering) around cations is not significantly affected by dispersion corrections. For instance, in the case of G-I salts, the solvation shell dimensions vary by at most ~0.08 Å across different dispersion models. Similarly, for G-II the variations are small (~0.05 Å).

**Table 2b**. Average cation-anion bond distance (First maxima in M-F PDF) for G-I and G-II salts across different dispersion models.

| | Ref | No-vdW | D2 | D3 | D3(BJ) | vdW-DF |
|---|---|---|---|---|---|---|
| **LiF** | 1.85 [37] | 1.84 | 1.86 | 1.82 | 1.84 | 1.86 |
| **NaF** | 2.3 [37] | 2.18 | 2.22 | 2.22 | 2.22 | 2.25 |
| **KF** | 2.6 [37] | 2.56 | 2.56 | 2.56 | 2.56 | 2.55 |
| **$BeF_2$** | 1.59[38] | 1.58 | 1.58 | 1.58 | 1.58 | 1.56 |
| **$MgF_2$** | 1.94[39]* | 1.9 | 1.93 | 1.93 | 1.93 | 1.96 |
| **$CaF_2$** | 2.23[39]* | 2.21 | 2.21 | 2.21 | 2.21 | 2.2 |

*Simulated with NNIP

Similarly, the first peak positions in the M-F RDF remain nearly identical across all models for G-I and G-II salts (Table 2b). Furthermore, the bond lengths observed in our study closely match previously reported values. This consistency suggests that M-F interactions in both Group-I and Group-II salts are primarily governed by columbic and attractive forces rather than dispersion interactions. This aligns with a previous study, speculating that charge interactions dominate molten salt structures.[40] Further, consistent M-M and F-F RDF features across different dispersion models reinforce that dispersion interactions play a minor role in these systems (numerical values for first minima in M-M and F-F RDFs are provided in Tables S2a and S2 b).

However, the notable differences appear in Be-Be RDF (Figure 2, lower panel). For instance, in the absence of dispersion, a small peak appears around 2.4 Å in the Be-Be RDF, which disappears when dispersion effects are included. To gain insight into this, we performed further analysis of bond length vs. bond angle (Figure S18) distributions, which showed a contraction of Be-Be distances leading to the observed pre-peak in Be-Be RDF in the case where no-vdW interactions were included. In this case, the inclusion of dispersion interactions corrected this by stabilizing the Be-Be distances and thus eliminating the pre-peak in the RDF.

For other G-II salts, such as $MgF_2$ and $CaF_2$, no such pre-peak is observed even in the absence of dispersion, suggesting that the effect of dispersion is most pronounced for small cations with higher charge density (related to smaller ionic radii of $Be^{2+}$(0.35Å) compare to $Mg^{2+}$ (0.72Å) and $Ca^{2+}$(0.99Å)) like $Be^{2+}$. This highlights the importance of dispersion corrections when modeling molten salts with high charge density species, ensuring a more accurate representation of their local structure, even under constant-density conditions.

#### *4.2.2. Coordination number*

Coordination number is a key structural parameter for understanding the local environment in molten salts and is known to affect salt properties. While the RDF explores the radial density of one atomic species relative to another, its integral (Eq. 6) yields the average coordination number ($CN_{avg}$) that explores the number of species in the vicinity of an atom, giving a precise idea of the local environment.

$$CN = 4\pi\rho \int_0^R r^2 g(r) dr \qquad (6)$$

Here, g(r) is the RDF, ρ is the density of the counter ion, and R is the distance of the first minimum of the RDF (provided in Table 2a). Table 3 and Figure S13 show the cation-anion $CN_{avg}$ for G-I and G-II molten fluoride salts.

**Table 3**. Cation-Anion coordination number as calculated from the integration of the PDF, up to the first minimum.

| CN | Ref. | No-vdW | D2 | D3 | D3(BJ) | vdW-DF |
|---|---|---|---|---|---|---|

| **LiF** | 3.7[37] | 4.36 | 4.62 | 4.18 | 4.36 | 4.23 |
|---|---|---|---|---|---|---|
| **NaF** | 4.1 [37] | 5.39 | 5.42 | 5.35 | 5.20 | 5.09 |
| **KF** | 4.9 [37] | 5.21 | 5.43 | 4.99 | 5.09 | 5.27 |
| **$BeF_2$** | 4[38] | 3.97 | 3.99 | 3.97 | 3.97 | 3.99 |
| **$MgF_2$** | 5.38*[39] | 5.21 | 5.29 | 4.97 | 5.10 | 5.1 |
| **$CaF_2$** | 7.03*[39] | 6.51 | 6.39 | 6.45 | 6.50 | 6.55 |

*Simulated values

The $CN_{avg}$ values computed across different dispersion models are first compared with previously reported results (Table 3). For G-I salts, the reported (Experimental) values ($CN_{Li-F}$ = 3.7 at 1123 K, $CN_{Na-F}$ = 4.1 at 1273 K, $CN_{K-F}$ = 4.9 at 1143 K)[37] are slightly higher than the computed $CN_{avg}$ across all the dispersion models. This discrepancy may arise due to differences in the radial cutoff used in experiments versus simulations. Additionally, the referenced $CN_{avg}$ values were determined at different temperatures and are often derived by fitting the total radial distribution function (RDF), which can introduce these differences.

For $BeF_2$, we obtained a $CN_{Be-F}$ value of approximately ~4 for all the dispersion methods, which is in good agreement with the experimental value of 4.[38] For $MgF_2$ and $CaF_2$, the experimental values of $CN_{avg}$ were unavailable. In these cases, we utilized $CN_{avg}$ derived from NNMD simulations of binary $MgF_2$-$CaF_2$ salt (NNIP trained on AIMD simulations with PBE functional with DFT-D3 dispersion), yielding $CN_{Mg-F}$ = 5.38 and $CN_{Ca-F}$ = 7.03 at T=1773K.[39] The computed values in this study are slightly lower, likely due to the temperature difference, as higher temperatures generally lead to lower coordination numbers. Nevertheless, these comparisons serve as meaningful benchmarks for assessing the performance of different dispersion models.

The sensitivity of $CN_{avg}$ to the choice of dispersion correction is further examined by measuring deviations relative to the dispersion-free (no-vdW) case. For G-I fluorides, LiF, $CN_{avg}$ remains close to 4, with a decrease of 4.1% using the D3 model and an increase of 6% under D2. The D3(BJ) models yield results closely aligned with the no-vdW case, and vdW-DF decreases CNavg by 3%. A similar trend is observed for NaF, where $CN_{avg}$ is 5.38 without dispersion. The D2 model slightly increases $CN_{avg}$ by 0.6%, while D3, D3(BJ), and vdW-DF reduce it by 0.7%, 3.49%, and 5.6%, respectively. In KF, dispersion corrections have a more pronounced effect: D3 decreases $CN_{avg}$ by 4%, and D3(BJ) by 2.3%, whereas vdW-DF and D2 increase it by 1.2% and 4.2%, respectively.

For G-II fluorides, different dispersion corrections also affect the $CN_{avg}$. In $BeF_2$, $CN_{avg}$ remains largely unaffected, with only a 0.6% increase under D2/vdW-DF. In $MgF_2$, the D3, D3(BJ), and vdW-DF decrease CNavg by 4.2%, 2.1%, and 2.1%, respectively, whereas D2 lead to slight increases of 1.5%. In $CaF_2$, semi-empirical dispersion corrections (D2, D3, and D3(BJ)) reduce $CN_{avg}$ by 0.6–1.9%, while vdW-DF increases it by 0.6%.

To investigate the root cause of the observed relative differences in the $CN_{avg}$ further, we explored the different metastable coordination numbers and their stability (Figure 3) in terms of their relative free energy. This analysis further gives us the idea of whether the dispersion corrections induce significant structural changes or if the variations in $CN_{avg}$ are minor and do not alter the underlying coordination number distribution. Here, we used a smooth function to explore different metastable coordination states using Eq 7[8]:

$$n = \sum_{i=1}^{N_F} \frac{1-\left(\frac{r_i}{r^{\dagger}}\right)^{12}}{1-\left(\frac{r_i}{r^{\dagger}}\right)^{24}} \tag{7}$$

Where, $r^{\dagger}$ is the first minimum of the cation-anion RDF (Table 2a), $r_i$ is the distance between the cation and anion, and $N_F$ is the number of fluorine atoms. The free energy profile (w(n)) of these coordination numbers is then calculated with the help of Eq. 8 in terms of their probability distribution $p$.

$$w(n) = K_B T \ln\left(p(n)\right) \tag{8}$$

Where $K_B$ is the Boltzmann constant, and *p(n)* is the probability of observing coordination numbers. The performance of the different dispersion models in capturing the metastable coordination states of the G-I and G-II molten salts exhibited clear trends as shown in Figure 3.

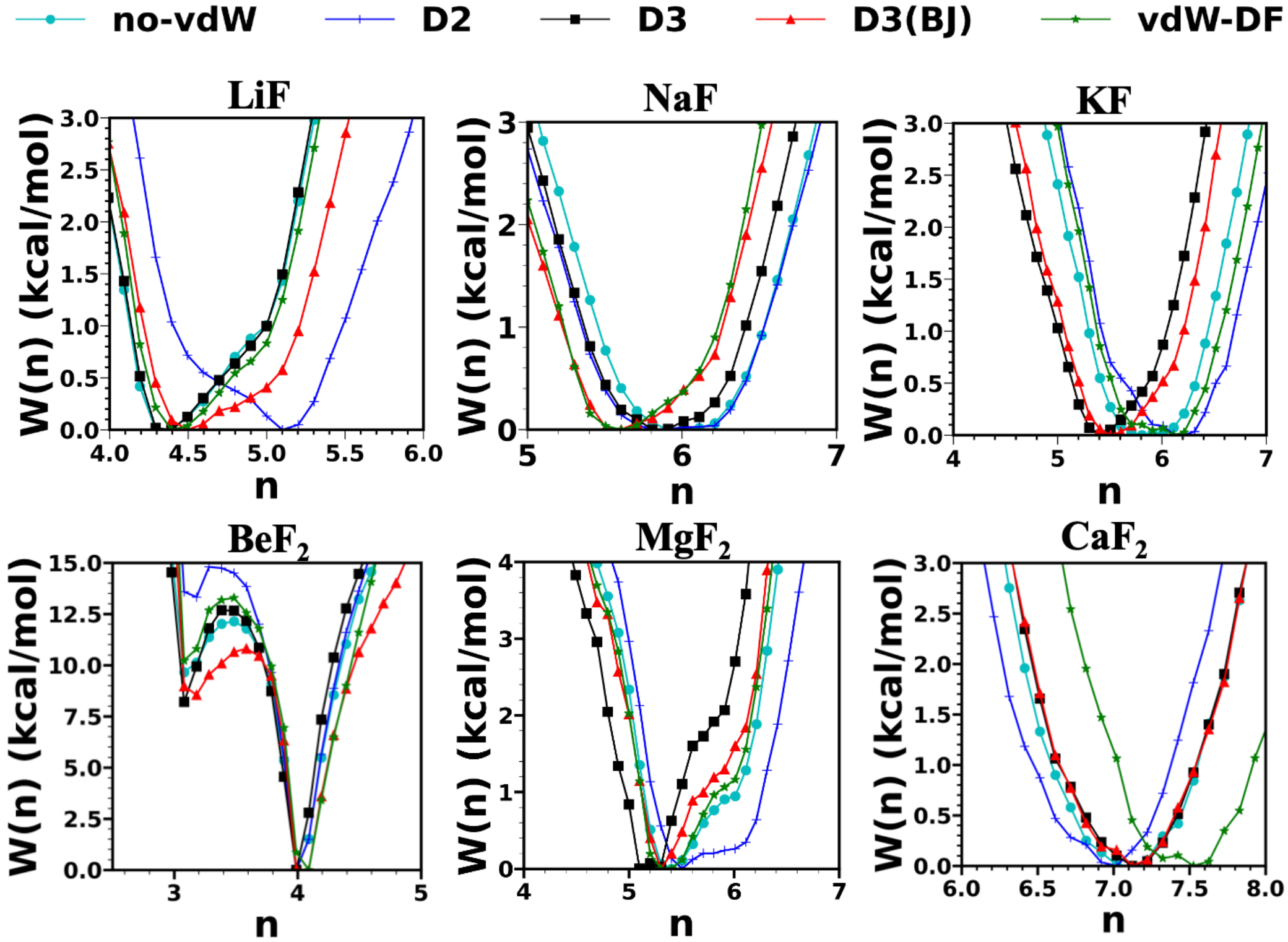

**Figure 3**. Free energy profiles display the fluoride coordination number distributions around different cations. These profiles were derived from Eqs (7-8) from MD trajectories.

For LiF, the D2 model predicts CN ~ 5 as the most stable state, whereas all the other methods converge at CN ~ 4.5. In NaF, the most stable CN varies between 5.5-6, depending on the model, with D2 and no-vdW prefer higher CN, and D3(BJ)/vdW-DF most stable around CN~5.5. Similarly, in KF, D3/D3(BJ) showed a preference towards lower CN (~5), whereas D2 and vdW-DF showed a preference for higher CN (~6).

In $BeF_2$, CN = 4 is consistently predicted as the most stable state by all methods. However, a minor amount CN = 3 is observed across all the dispersion models. For $MgF_2$, all models predict CN = 5 as the most stable state, with minor contributions from CN = 6, which is slightly more stabilized in the case of D2. For $CaF_2$, all models agree that CN = 7 is the most stable, with vdW-DF shifting slightly toward higher CNs, indicating its preference for an overcrowded coordination environment.

Our analysis demonstrates that while the RDF profiles remain largely invariant across different dispersion corrections, the $CN_{avg}$ derived from these RDFs exhibits measurable sensitivity. Semi-empirical dispersion models such as D2 and D3/D3(BJ) exhibit opposing effects: D3 and D3(BJ) tend to destabilize higher coordination environments, thereby reducing $CN_{avg}$ relative to the no-vdW reference. In contrast, D2 and vdW-DF tend to stabilize higher coordination states, resulting in an increase in $CN_{avg}$. The behavior of vdW-DF is more nuanced: in some systems (e.g., KF, $CaF_2$), it slightly increases CNavg, while in others (e.g., LiF, NaF, $MgF_2$), it causes a modest decrease. This inconsistency suggests that vdW-DF introduces system-specific variations, likely due to its nonlocal treatment of dispersion forces.

Although these shifts in coordination number may appear subtle, they have significant implications for the prediction of thermophysical and spectroscopic properties, including vibrational dynamics and ionic transport. Notably, our observations are consistent with recent work by Rajni et al., who showed that the choice of dispersion correction can substantially affect the Raman spectra of molten $AlCl_3$[41], most likely due to alterations in the underlying coordination number distributions. These findings underscore the importance of carefully selecting and benchmarking dispersion models against experimental or high-level theoretical data to ensure the accuracy and reliability of molten salt simulations.

#### *4.2.3. Angular distribution function*

While RDF and CN give us an idea of the density of a species around another species angular distribution function (ADF) helps one to understand the atomic connectivity/topology of the local solvation shell. While the ADF for the anion-cation-anion (F-M-F) (Figure S17) configurations provides insights into the geometric arrangement surrounding the cation, the cation-anion-cation (M-F-M) ADF (Figure 4) explores the inter-cation connectivity. The ADFs across different dispersions and species are computed using the radial cutoff provided in Table 2a and normalized in a way that the area under the ADF remains unity.

For G-I salts, the $\theta_{F-M-F}$ ADF (Figure S17, upper panel) plots exhibit a characteristic double-peak structure, which is consistently captured across all the models, with one peak between 80°–100° and another around 150°. This distribution reflects the flexibility of the solvation shell, where multiple angular preferences exist due to the possibility of different local coordination environments. The presence of these peaks is consistent with the CN analysis, which shows that G-I salts can adopt multiple coordination states, leading to variations in solvation structure.

G-II salts such as $MgF_2$ and $CaF_2$ exhibit $\theta_{F-M-F}$ ADF distributions that resemble those of G-I molten salts, with broader peaks centered around 90° and 150°. This suggests that, similar to G-I, $Mg^{2+}$, and $Ca^{2+}$, experience greater solvation shell flexibility due to the coexistence of five- and six-fold coordination states. However, in contrast to other salts, $BeF_2$ exhibits a single peak around ~110°, confirming the tetrahedral coordination of $BeF_4^{2-}$ units, which is consistent with the $\theta_{F-Be-F}$ ADF of FLiBe.[42]

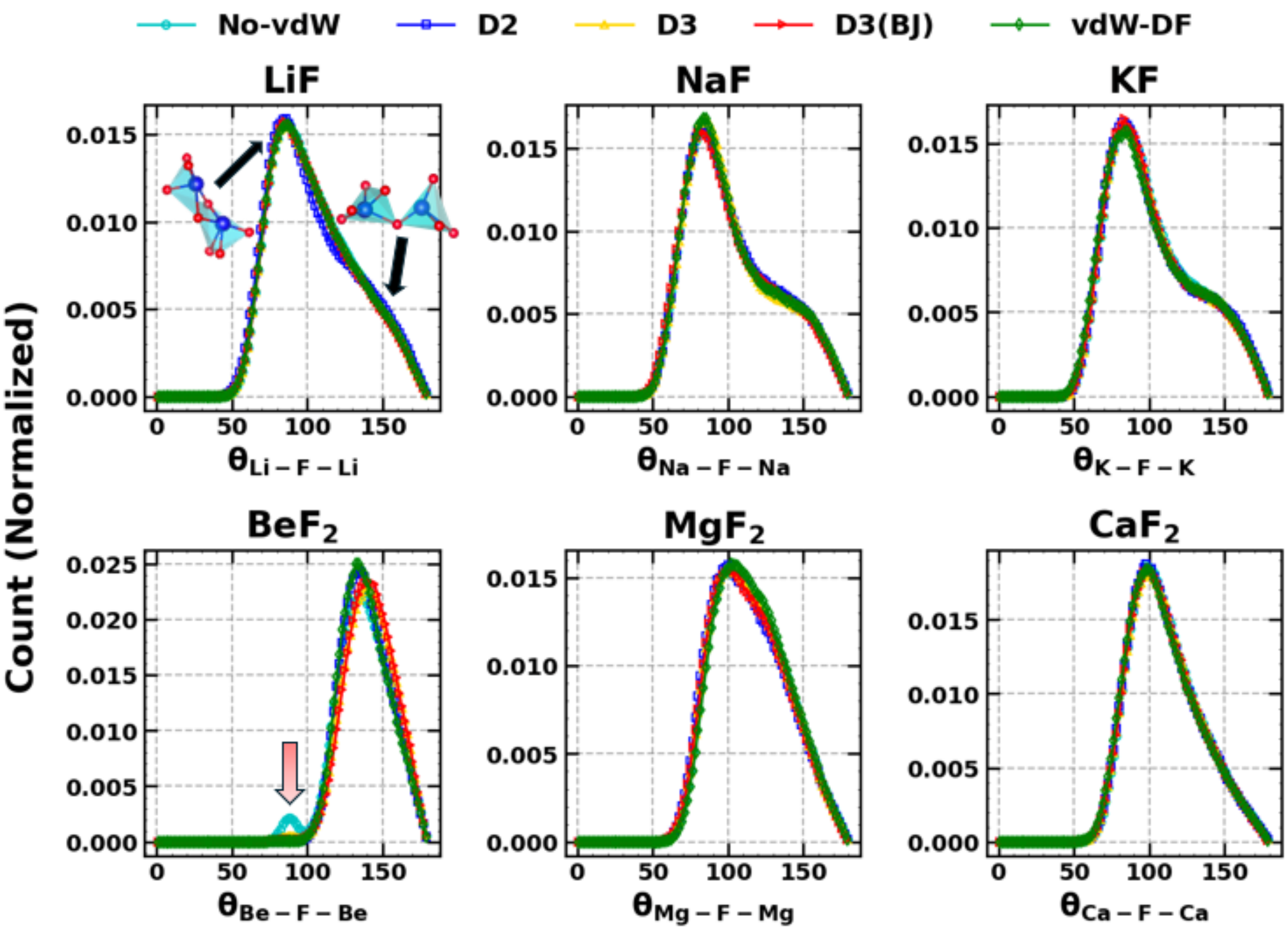

**Figure 4**. $\theta_{M-F-M}$ ADF computed with different dispersions for G-I and G-II molten salts. The peak around 80° in $BeF_2$ is pointed out with a shaded red arrow. An example of the corner and edge-sharing complex is also shown in the LiF ADF.

The $\theta_{M-F-M}$ ADF (Figure 4) shows consistent results across different dispersion models for both G-I and G-II salts, with the primary distinction being observed in $BeF_2$. For $BeF_2$, all dispersion models yield a sharp peak in the $\theta_{Be-F-Be}$ ADF around ~140°, consistent with previous observations for FliBe melts. However, in the absence of vdW, an additional small peak emerges at lower angles (~80-90°), which is not present when dispersion interactions are included. This is consistent with the previously noted contraction of Be-Be distances in the absence of dispersion interaction (bond contraction leads to angle contraction, Figure S18). Moreover, a sharp peak around ~140° can be seen for all the models for $BeF_2$.

Overall, the ADF analysis shows that the solvation structure and atomic connectivity remain largely insensitive to the choice of dispersion model, with all functionals capturing the same fundamental features. This further emphasizes our previous speculation of predominant columbic interaction in determining salt structure. However, in $BeF_2$, dispersion corrections become important, and inclusion of dispersion ensures a more relaxed solvation environment without altering the overall topology.

### *4.2.4. Binding Free Energy*

In general, ion-ion dynamics is crucial for understanding the thermodynamic and transport properties of molten salts.[43,44] This ion-ion dynamics is strongly influenced by ion-ion binding energy. Typically, the binding energy of two ions can be calculated from their respective potential mean force (PMF).[45] In physical terms, PMF ($w_{ij}^{eff}$) explores the energy barrier for atom exchange across the solvation shell. This quantity can be calculated from the ion-ion RDF with the help of Eq. 9,[46]

$$\beta w_{ij}^{eff} = -\ln\left(g_{ij}(r)\right) \tag{9}$$

where $\beta = \frac{1}{K_B T}$ and $g_{ij}(r)$ s RDF. The cation-anion barrier (Figure S20) is calculated as the difference between the first maxima and the minima of the PMF (Figure S19) for Group I and Group II molten salts, computed across different dispersion models. The values of the barrier height are provided in Table 4.

Table 4. Barrier height calculated from Cation-Anion PMF. All the values are in $\beta^{-1}$ units.

| Barrier | No-vdW | D2 | D3 | D3(BJ) | vdW-DF |
|---|---|---|---|---|---|
| **LiF** | 2.045 | 1.998 | 1.98 | 2.022 | 1.975 |
| **NaF** | 2.017 | 2.051 | 2.009 | 2.025 | 1.979 |
| **KF** | 1.89 | 1.941 | 1.865 | 1.932 | 1.887 |
| **$BeF_2$** | 6.642 | 7.236 | 6.666 | 6.283 | 7.018 |
| **$MgF_2$** | 2.48 | 2.392 | 2.450 | 2.465 | 2.44 |
| **$CaF_2$** | 1.953 | 1.928 | 1.948 | 1.937 | 1.939 |

Among the studied salts, the exchange barrier remains consistently around ~2$\beta^{-1}$ units, except for $BeF_2$, which exhibits a significantly higher barrier (~6–7$\beta^{-1}$ units) due to the strong coulombic attraction between $Be^{2+}$ and $F^-$, driven by the high charge density of $Be^{2+}$.[40] While dispersion corrections introduce minor variations (±1.5%) for most salts, $BeF_2$ shows greater sensitivity (±5%), with D2 predicting the highest barrier (7.236$\beta^{-1}$) and D3(BJ) the lowest (6.283$\beta^{-1}$). This sensitivity arises from the interplay of Coulombic interactions and $Be^{2+}$ polarization effects, which enhance dispersion contributions.

The elevated exchange barrier aligns with $BeF_2$'s rigid, network-like solvation structure mentioned in past literature,[7,35,47] which again confirmed by CN and ADF results, indicate stable tetrahedral $BeF_4^{2-}$ coordination with minimal solvation fluctuations. These findings highlight the necessity of selecting appropriate dispersion models for highly polarizing cations like $Be^{2+}$, offering valuable insights for optimizing molten salts in energy storage and electrochemical applications.

**4.3. Transport Properties: Self-diffusion coefficients**

Accurate prediction of transport properties such as viscosity and diffusivity are critical for system design. While longer (~ns) and larger box simulations are required for calculating viscosity, relatively short timescale (~ps) simulations can be used to evaluate diffusion coefficients. In the present study, we focus on evaluating the self-diffusion coefficients due to the high computational cost associated with AIMD, particularly when screening multiple G-I and G-II salt compositions across various dispersion models.

The diffusion coefficients were computed using the Einstein method, as defined in Eq. 10a and Eq. 10b, by fitting the linear regime of the mean squared displacement (MSD) curves.

$$MSD = \frac{1}{N}\frac{1}{n_t}\sum_{j=0}^{N}\left(r_i\left(t_{0j}+t\right)-r_{i\left(t_{0j}\right)}\right)^2 \tag{10a}$$

$$D = \frac{1}{6}\,lim_{\{t\rightarrow\infty\}}\frac{d}{dt}(MSD) \tag{10b}$$

To improve statistical reliability, we employed a block-averaging method[48], wherein each trajectory was partitioned into equal time segments and diffusion was calculated for each block independently. The final reported value is the average across blocks, and the associated standard error provides an estimate of the reliability of the observable.

To correct for finite-size effects arising from the relatively small AIMD simulation cells, we applied the Yeh–Hummer[49] correction to all computed diffusion coefficients. The details of this method can be found in the supplementary material. Corrected values are reported throughout the analysis.

Figure 5 and Figure 6 show the computed temperature-dependent diffusion coefficients for G-I and G-II salts, respectively, across five dispersion correction models (no-vdW, D2, D3, D3(BJ), and vdW-DF). For most systems, the diffusion coefficients exhibit consistent Arrhenius behavior, with values ranging from $10^{-5}$ to $10^{-4}$ $cm^2/s$ depending on the salt and temperature. Relative standard errors (calculated based on the block averaging) for G-I salts are generally below 10%, indicating that 30–40 ps trajectories yield converged and reliable diffusion estimates. This is further supported by previous studies showing that 20–30 ps AIMD simulations are sufficient for accurately calculating diffusion coefficients in high-temperature molten salts[50,51]. The same applies to G-II salts with the exception of $BeF_2$, which is discussed separately below.

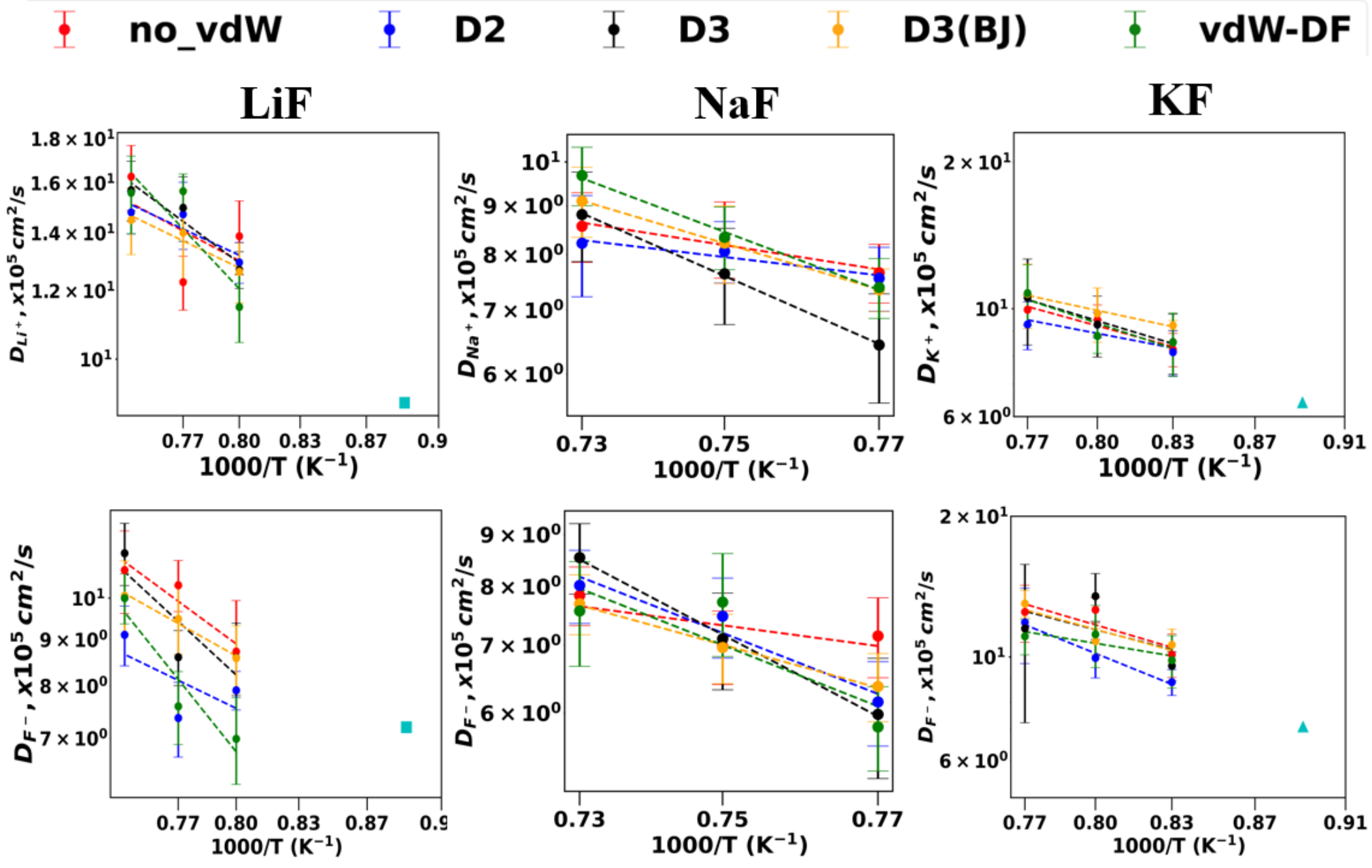

**Figure 5**. Diffusion coefficients for G-I molten salts were calculated across different temperatures and dispersion models with Einstein's method (90% CI). For LiF, the experimental values are marked with squares [52]. Simulated values for KF are marked as triangles [52]

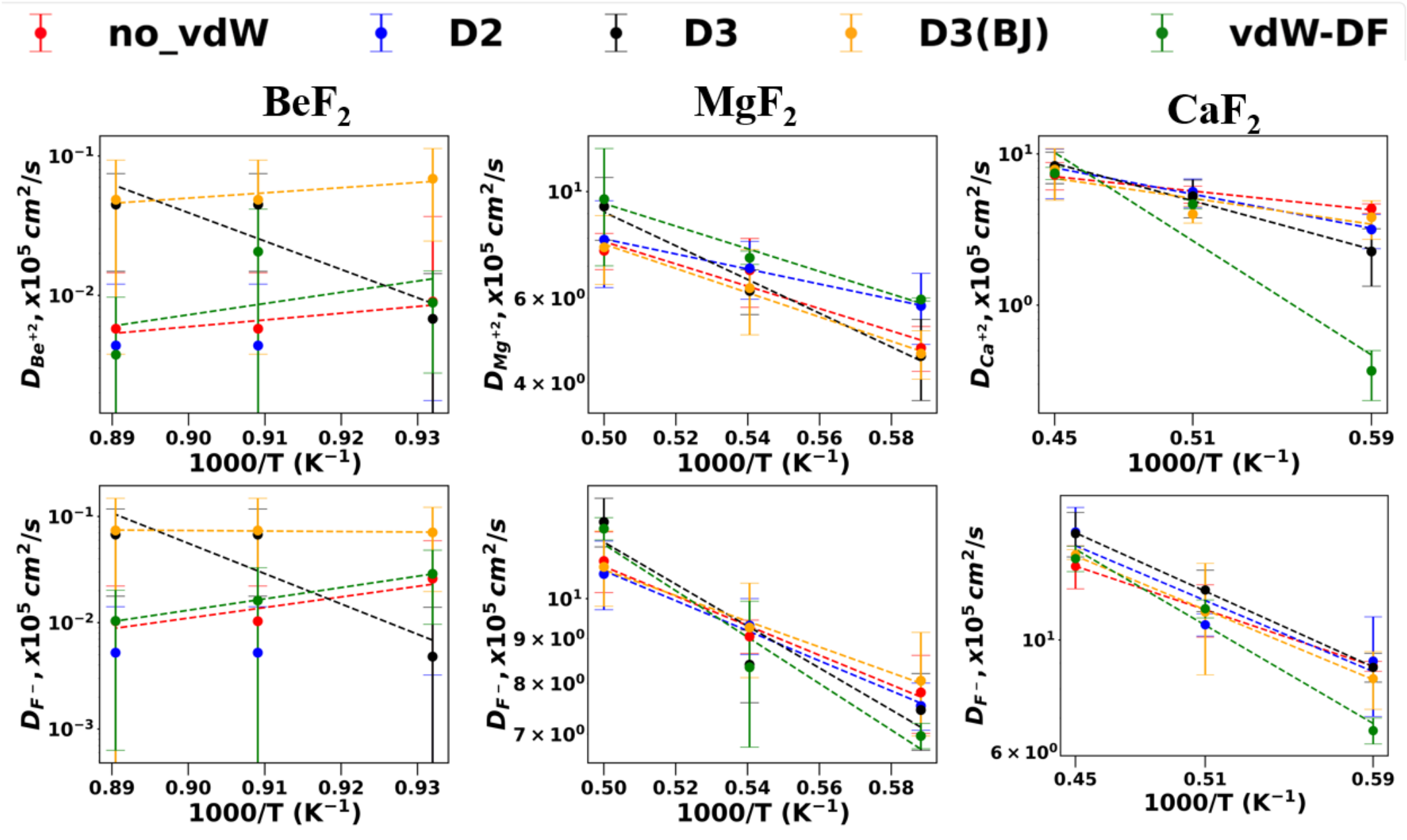

**Figure 6**. Diffusion coefficients for G-II molten salts were calculated across different temperatures and dispersion models with Einstein's method (90% CI).

To further validate our results, we compare the computed diffusion coefficients with available experimental/simulated data. For LiF, for all methods, the calculated diffusion coefficients for Li+ and F- at 1250–1350 K are in good agreement with experimental values at a comparable temperature (1123 K), which report $7.2 \times 10^{-5}$ cm²/s for F and $8.9 \times 10^{-5}$ cm²/s for Li.[52] Similarly, the obtained diffusion coefficients for K and F for all methods for KF are in good agreement with the previously reported simulation value of $6.42 \times 10^{-5}$ cm²/s and $7.1 \times 10^{-5}$ cm²/s for K and F, respectively, at 1123 K.[52] For NaF, although direct experimental diffusion data are unavailable, the predicted diffusion coefficients of $Na^+$ and $F^-$ are comparable to reported values for these ions in FLiNaK[53] mixtures further support their plausibility. No experimental diffusion data are currently available for the G-II fluorides, but our predicted values follow consistent Arrhenius trends and fall within physically reasonable ranges of the similar salt systems.[39,54] These comparisons demonstrate that the computational approach employed in this study yields reliable transport properties for alkali halide molten salts. While dispersion interactions have been shown to significantly impact diffusion in water, increasing it by nearly two to threefold[55], the same is not observed in G-I and G-II molten salts.[36]

Across G-I and G-II salts (excluding $BeF_2$), diffusion coefficients vary only modestly between dispersion models, typically within 10–20%. Moreover, for most cases, given the uncertainty associated with the calculations of diffusion coefficients, their values remain largely insensitive to the choice of dispersion correction. This observation can be attributed to the previously observed ion exchange barrier (Table 4)-similar barriers across different dispersions yield similar diffusion coefficients.

An exception to the general trend is observed for $CaF_2$ at 1700 K, where both $Ca^{2+}$ and $F^-$ diffusion coefficients are systematically lower when using the vdW-DF functional. This deviation is attributed to vdW-DF's overestimation of the liquid density near the melting point, as evidenced by the strongly negative excess pressures at this temperature (Figure 1, Figure S7). Notably, only vdW-DF exhibits a pronounced temperature dependence in the density error for $CaF_2$, with the overestimation decreasing significantly at higher temperatures. This behavior is not observed in the other dispersion models (D2, D3, D3(BJ)), which yield smaller and relatively constant density deviations across temperatures. The nonlocal formulation of vdW-DF likely enhances sensitivity to structural compaction effects near the melting regime, leading to reduced ionic mobility. For other salts, simulated well above their melting points, both density and diffusion coefficients remain largely unaffected by the choice of dispersion correction.

$BeF_2$ represents a distinct case. For all dispersion models, the computed diffusion coefficients of $Be^{2+}$ and $F^-$ are significantly lower than those observed in other systems, by approximately an order of magnitude, falling into the range of $10^{-7}$ to $10^{-8}$ $cm^2/s$ at typical operating temperatures. This slow diffusion is consistent with prior reports using polarizable force fields ($D_{F^-}$ ~$10^{-8}$ $cm^2/s$)[7] and reflects the formation of persistent, extended Be–F polymeric networks in the melt. Such networks introduce structural rigidity and hinder ion mobility, leading to long correlation times and high statistical noise in MSD calculations.

Indeed, for $BeF_2$, the diffusion coefficients exhibited larger relative uncertainties (up to 15–20%) compared to other salts. This increased variability arises from the much slower ion dynamics in $BeF_2$, where the formation of extended Be–F polymeric networks significantly reduces ionic mobility. As a result, the timescales accessible to AIMD may not fully capture long-time diffusive behavior, leading to higher uncertainty and potential deviation from ideal Arrhenius behavior. While the diffusion coefficients extracted for $BeF_2$ provide a qualitative estimate of ionic mobility, they should be interpreted with caution due to the inherent limitations of sampling slow dynamics within accessible AIMD timescales.

Dispersion corrections had a more pronounced effect on $BeF_2$ than on other systems. Notably, the D3(BJ) correction consistently yielded higher diffusivity than other models, which is consistent with its previously observed tendency to reduce the Be–F exchange barrier (Table 4). In contrast, simulations without dispersion corrections yielded the lowest diffusion coefficients. These results underscore the importance of selecting an appropriate dispersion model when modeling systems with small, highly charged, and highly polarizable cations like $Be^{2+}$. Overall, our findings suggest that at fixed density, diffusion coefficients in molten salts are largely insensitive to the choice of dispersion model—except in structurally complex systems such as $BeF_2$.

## 5. CONCLUSION

Our study highlights the role of the selection of appropriate dispersion interaction schemes in molten salt simulations. In this study, we explored different dispersion methods for salts (G-I :(LiF, NaF, KF) and G-II:($BeF_2$, $MgF_2$, $CaF_2$)) and studied the effect of the chosen dispersion scheme on the predicted density, local structure (RDF, CN) as well as transport and thermophysical properties of each salt. To evaluate the prediction accuracy of each dispersion scheme, we used experimental data available in different literature. For most salts, simulations at constant density (here, experimental density) yield consistent structural and transport properties regardless of the dispersion model. However, we report that density predictions are highly sensitive to the chosen dispersion model, with errors reaching up to 40% (NaF with vdW-DF). Notably, our simulations indicate that semiempirical dispersions are often better for density prediction compared to the nonlocal vdW-DF functional. Moreover, our study indicates that although the key features of ADF and RFDs remain consistent across all dispersion models for most of the salts, the CN is significantly affected by the choice of dispersion functionals.

Unlike other salts studied in this work, $BeF_2$, in particular, displays significant structural and dynamic discrepancies in the absence of dispersion corrections, emphasizing the need for careful dispersion model selection. The pronounced effects observed in $BeF_2$ lead us to consider whether the salts with intermediate-range ordering, such as $UF_4$ and $ZrF_4$, may exhibit similar behavior, necessitating further investigation.

To provide practical guidance for future studies, we summarize the optimal dispersion models for predicting density, diffusion coefficients, reaction barriers, and structural properties in Table 5.

**Table 5:** Recommended dispersion models for molten fluoride salts

| **Salt** | **Density** | **Diffusion** | **Barrier** | **RDF** | **ADF** | **CN** |
|---|---|---|---|---|---|---|
| **LiF** | vdW-DF | vdW-DF | vdW-DF | vdW-DF | vdW-DF | vdW-DF |
| **NaF** | D3(BJ) | D3(BJ) | D3(BJ) | D3(BJ) | D3(BJ) | D3(BJ) |
| **KF** | D3/D3(BJ) | D3/D3(BJ) | D3/D3(BJ) | D3/D3(BJ) | D3/D3(BJ) | D3/D3(BJ) |
| **$BeF_2$** | D3 | D3 | D3(BJ) | D3/D3(BJ) | D3/D3(BJ) | D3/D3(BJ) |
| **$MgF_2$** | D3 | D3 | D3 | D3 | D3 | D3 |
| **$CaF_2$** | D2 | D2 | D2 | D2 | D2 | D2 |

Overall, this study serves as a practical guide for selecting appropriate dispersion models for simulating LiF, NaF, KF, $BeF_2$, $MgF_2$, and $CaF_2$ molten salts in accurately predicting structure, thermodynamic, and transport properties of these systems across a wide range of temperatures. The workflow presented in this work can be further expanded to explore appropriate dispersion methods for other molten salt systems of interest. Moreover, knowledge about the accurate dispersion model for each system will help us to develop a more accurate NNIP. This, however, will be the focus of our future work.

**Data availability**

The datasets (MD trajectories and other output files) used and/or analyzed during the current study are available from the corresponding authors upon reasonable request.

## ACKNOWLEDGEMENTS

This work was supported by DOE-NE's Nuclear Energy University Program (NEUP) under Award DE-NE0009204. A part of the computational resources was provided by the Massachusetts Green High-performance Computing Cluster (MGHPCC). This research also used resources of the National Energy Research Scientific Computing Center, a DOE Office of Science User Facility supported by the Office of Science of the U.S. Department of Energy under Contract No. DE-AC02-05CH11231, award ASCR-ERCAP0022362 and BES-ERCAP0022445.